\documentclass[namedreferences]{solarphysics}
\usepackage[optionalrh]{spr-sola-addons} 
\usepackage{graphicx}        
\usepackage{amssymb}        
\usepackage[usenames,dvipsnames]{color}

\usepackage{hyperref}             

\begin{document}

\begin{article}

\begin{opening}

\title{Statistical Survey of Type III Radio Bursts at Long Wavelengths Observed by
the \textit{Solar TErrestrial RElations Observatory} (STEREO)/\textit{Waves} Instruments:
Radio Flux Density Variations with Frequency}

\author{V.~\surname{Krupar}$^{1}$\sep
        M.~\surname{Maksimovic}$^{2}$\sep
        O.~\surname{Santolik}$^{1,3}$\sep
        E.~P.~\surname{Kontar}$^{4}$\sep
				B.~\surname{Cecconi}$^{2}$\sep
        S.~\surname{Hoang}$^{2}$\sep
				O.~\surname{Kruparova}$^{1}$\sep
				J.~\surname{Soucek}$^{1}$\sep
				H.~\surname{Reid}$^{4}$\sep
				A.~\surname{Zaslavsky}$^{2}$
       }
\runningauthor{V. Krupar \textit{et al.}}
\runningtitle{Type III Radio Bursts}

   \institute{$^{1}$ Institute of Atmospheric Physics ASCR, Bocni II 1403, Prague 141 31, Czech Republic\\
                     email: \url{vk@ufa.cas.cz}\\
                    $^{2}$ LESIA, UMR CNRS 8109, Observatoire de Paris, Meudon 92195, France\\
              $^{3}$ Faculty of Mathematics and Physics, Charles University, Ke Karlovu 3, Prague 121 16, Czech Republic\\
							                    $^{4}$ Department of Physics and Astronomy, University of Glasgow, Glasgow G12 8QQ, UK}

\begin{abstract}
We have performed a statistical study of $152$ Type III radio bursts observed
by \textit{Solar TErrestrial RElations Observatory} (STEREO)/\textit{Waves} between May 2007 and February 2013.
We have investigated the flux density between $125$~kHz and $16$~MHz. Both high- and low-frequency cutoffs
have been observed in $60$\,\% of events suggesting an important role of propagation.
As already reported by previous authors, we observed that the
maximum flux density occurs at $1$~MHz on both spacecraft. We have developed a simplified analytical model of the flux density as a
function of radial distance and compared it to  the STEREO/\textit{Waves} data.

\end{abstract}
\keywords{Solar radio emissions, Plasma radiation}
\end{opening}

\section{Introduction} 
      \label{S-intro}

Type III radio bursts are consequence of suprathermal electrons accelerated during solar flares \cite{1950AuSRA...3..541W}.
The Type III-generating electron beam propagates outward from the Sun along an open magnetic field line in the interplanetary (IP) medium
with speed ranging from $\approx0.3$c to $\approx0.05$c \cite{1987A&A...173..366D,1999SoPh..184..353M,2000GMS...119..115D}. 
As the electron beam produces the bump-on-tail instability, it locally excites intense
Langmuir waves at the local electron plasma frequency [$f_{\rm{pe}}$]. These waves then convert to electromagnetic emission 
via a series of non-linear processes which are still debated and require better understanding
\cite{1980SSRv...26....3M,1998SoPh..181..429R}.

The generation of Langmuir waves by the beam also produces back-reaction
on the local distribution of electrons,
which makes the problem of electron propagation inherently non-linear.
The characteristic scales of electron-beam propagation ($\approx1$~AU) require
additional simplifications such
as quasi-linear treatment \cite{ISI:A1962WS88500022,ISI:A1962WS88400006,1962PhRv..125..804P}.
The relatively low level of the waves is normally justified
by the observations of the mean electric fields associated with Type III
exciters.
The numerical simulations of transport and generation as well as
re-absorption of Langmuir waves show
that the electron beam can propagate over large distances with
relatively weak losses while at the same
time producing a high level of plasma waves locally
\cite{1977SoPh...55..211M}.
The important aspect of beam--Langmuir wave evolution is the sensitivity
to plasma--density fluctuations in the solar corona and the heliosphere. The
density fluctuations can effectively
influence the distribution of waves, which in turn affects the electron
distribution \cite{1979ApJ...233..998S,1986A&A...163..229M}.
Numerical simulations \cite{2001A&A...375..629K} also reveal that
the plasma inhomogeneities change the spatial distribution
of the waves, while the local electron distribution is weakly affected.
However, the effects of density fluctuations and large--scale
density gradient in the solar plasma becomes noticeable at large
distances \cite{2009ApJ...695L.140K}, so that
initially injected power-law spectrum becomes a broken one. Overall, the
rate at which plasma waves are induced by an unstable electron
beam is reduced by background density fluctuations, most acutely when
fluctuations have large amplitudes or small wavelengths.
Numerical simulations \cite{2010ApJ...721..864R} further show a
direct correlation between the spectrum of the double
power-law below the break energy and the turbulent intensity of the
background plasma.

Langmuir waves can be converted via the plasma emission process \cite{1958SvA.....2..653G} into electromagnetic radiation: Type III radio bursts either
at $f_{\rm{pe}}$ [the fundamental
[\textit{F}] component],
and/or at $2f_{\rm{pe}}$ [the harmonic
[\textit{H}] component].
Although this mechanism has been extensively investigated, the conversion itself still remains under debate.
Type III radio bursts can be observed from metric to kilometric wavelengths \cite{2000ApJ...530.1049R}.
From metric to decametric wavelengths we can usually distinguish the \textit{F}- and
\textit{H}-components while 
the \textit{F}-component is often more intense than the \textit{H}-component.
At kilometric wavelengths it is problematic to determine the observed component.
\inlinecite{1980ApJ...236..696K} has developed a method for determination of the
component for cases when Type III triggering electron beams intersect the spacecraft.

Propagation of Type III radio bursts from
the source to the spacecraft is affected both by refraction in density gradients and scattering by inhomogeneities in the solar wind
ranging on all scales from $\approx100$~km to $\approx0.3$~AU.
These effects result in the shifted position of a source location,
enlargement of the source size, and decrease of the flux density and degree of polarization \cite{2000GMS...119..115D}.
Better understanding of proper images of radio sources may shed light on physical processes along the path
of the exciter electrons in the IP medium.

While previous solar spacecraft used for Type III radio-burst investigation were
all spin-stabilized
(\textit{International Sun/Earth Explorer 3},
\textit{Ulysses},
\textit{Wind},
\textit{etc.}), \textit{Solar TErrestrial RElations Observatory}
(STEREO; \opencite{2008SSRv..136....5K}) is the first three-axis stabilized solar mission
to observe these waves.
With the two identical STEREO spacecraft and their
suite of state-of-the-art instruments, we can study mechanisms
and sites of solar energetic-particle acceleration in the low corona and the IP medium.
The STEREO/\textit{Waves} instrument enables us to measure radio flux density from
decametric to kilometric wavelengths.
Therefore we can investigate global distributions of radio sources in the solar wind
when an appropriate electron-density model is used.

In this article,
we present statistical results on the frequency spectra of Type III radio bursts at long (\textit{i.e.}
from dekametric to kilometric)
wavelengths observed by STEREO/\textit{Waves}.
This is the first of two linked articles
that summarize our findings on statistical properties of Type III radio bursts.
In this article we focus on a flux density and its relation to electron beams producing these bursts.
Goniopolarimetric (GP; also referred to as direction-finding) results will
be discussed in the second article \cite{krupar2014_2}.

In Section~\ref{S-data} we describe the instrumentation and data processing.
In Section~\ref{S-Results} we analyze two Type III radio bursts and describe our statistical data set. Next we present a statistical examination of Type III radio
bursts with focus on the frequency spectra and the model of the flux density as a function of the radial distance. In Section~\ref{S-Conclusion} we summarize our findings and we make concluding remarks.

\section{Instrumentation and Methodology} 
      \label{S-data}

\subsection{The STEREO/\textit{Waves} Instrument} 
      \label{S-stereo}

STEREO provides us with first stereoscopic measurements of the solar phenomena using identical
instruments onboard \cite{2008SSRv..136...45B}. The two STEREO
spacecraft are three-axis stabilized and carry
the STEREO/\textit{Waves} instruments, which
measure electric-field fluctuations
between 2.5~kHz and 32.025~MHz \cite{2008SSRv..136..487B,2008SSRv..136..529B}.
The three monopole antenna elements (six meters long),
made from \textit{beryllium--copper}, are used by
STEREO/\textit{Waves}
to measure electric-field fluctuations \cite{2008SSRv..136..529B}.
Although we use three orthogonal antennas, their effective antenna directions [$\zeta_{\rm{eff}}$
and $\xi_{\rm{eff}}$] and lengths
[$l_{\rm{eff}}$]
are different from the physical ones due to their electrical coupling with
the spacecraft body.

For our survey we have used data from the \textit{High Frequency Receiver}
(HFR: a part of the STEREO/\textit{Waves} instrument; \opencite{2008SSRv..136..487B}),
which is a dual-channel receiver (connected to two antennas at one time) operating in the
frequency range $125$~kHz~--~$16.025$~MHz with a $25$ kHz effective bandwidth.
HFR has instantaneous GP capabilities between $125$~kHz and $1975$~kHz allowing us to retrieve the direction of arrival of an incoming electromagnetic
wave, its flux, and its polarization properties \cite{2008SSRv..136..549C}.
HFR consists of two receivers: HFR1 ($125$~kHz~\---~$1975$~kHz, $38$ frequency channels) and HFR2 ($2025$~kHz~\---~$16.025$~MHz, $281$ frequency channels).
HFR1 provides us with auto- and cross-correlations on all antennas (three monopoles), while HFR2 has
retrieved only two auto-correlations
(one monopole and one dipole) for most of the time since May 2007.

Using a semiempirical model of electron density [$n_{\rm{e}}$] in the solar corona and IP medium \cite{1999ApJ...523..812S} we can assign particular frequencies to radial distances from the Sun.
Consequently STEREO/\textit{Waves}/HFR allows us to investigate GP properties of radio
sources located between $4$~$\rm{R}_{\odot}$ and $40$~$\rm{R}_{\odot}$ while the intensity is measured up to $1$~$\rm{R}_{\odot}$ above the Sun's surface (Figure \ref{sittler}).

For converting the voltage power spectral density
at the terminals of the antennas [$V^2$:
$\rm{V^2Hz^{-1}}$]
into physical units we have applied the method described by \inlinecite{2011RaSc...46.2008Z}.
In order to compute the flux density [$S$:
$\rm{W m^{-2} Hz^{-1}}$] as below, we need to know the effective antenna lengths and receiver properties.

\begin{equation}  \label{eq:s}
S~=~\frac{V^2}{Z_0\Gamma^2l_{\rm{eff}}^2},
\end{equation}

\begin{equation}  \label{eq:gamma}
\Gamma~=~\left|\frac{C_{\rm{a}}}{C_{\rm{a}} + C_{\rm{s}}}\right|,
\end{equation}
where $Z_0=\sqrt{\mu_0/\epsilon_0} \approx 120 \pi$ denotes the impedance of vacuum.
The $\Gamma$-coefficient is the gain factor of the receiver-antenna
system and is composed from the antenna capacitance [$C_{\rm{a}}$] and
the stray capacitance [$C_{\rm{s}}$].

The model of the galactic background brightness [$B_{\rm{model}}$],
as a nearly stable
isotropic source, allows us to determine reduced effective antenna lengths
[$\Gamma l_{\rm{eff}}$] according to \inlinecite{2011RaSc...46.2008Z}:
\begin{equation}  \label{eq:gammaleff}
\Gamma l_{\rm{eff}}~=~\left( \frac{3}{4\pi Z_0} \frac{V^2-V^2_{\rm{noise}}}{B_{\rm{model}}} \right)^{1/2},
\end{equation}
\begin{equation}\label{eq:galaxy}
B_{\rm{model}}~=~B_0f^{-0.76}_{\rm{MHz}}e^{-\tau},
\end{equation}
\begin{equation}\label{eq:galaxy2}
\tau~=~3.28f^{-0.64}_{\rm{MHz}}
\end{equation}	
where $B_0~=~1.38\times 10^{-19}$~$\rm{Wm^{-2}Hz^{-1}sr^{-1}}$ and $f_{\rm{MHz}}$ is the frequency in MHz \cite{1978ApJ...221..114N}.
The $V^2_{\rm{noise}}$ represents a square noise generated by the receiver itself and hence should be subtracted
before the antenna calibration. We have adapted $V^2_{\rm{noise}}$ values from \inlinecite{2011RaSc...46.2008Z}:

\begin{itemize}
	\item $V^2_{\rm{noise}}= -160.9$~dB for STEREO-A $_{\rm{channel1}}$
	\item $V^2_{\rm{noise}}= -162.8$~dB for STEREO-A $_{\rm{channel2}}$
	\item $V^2_{\rm{noise}}= -162.4$~dB for STEREO-B $_{\rm{channel1}}$
	\item $V^2_{\rm{noise}}= -165.9$~dB for STEREO-B $_{\rm{channel2}}$
\end{itemize}

\begin{figure}    
   \centerline{\includegraphics[width=0.7\textwidth,clip=,angle=90]{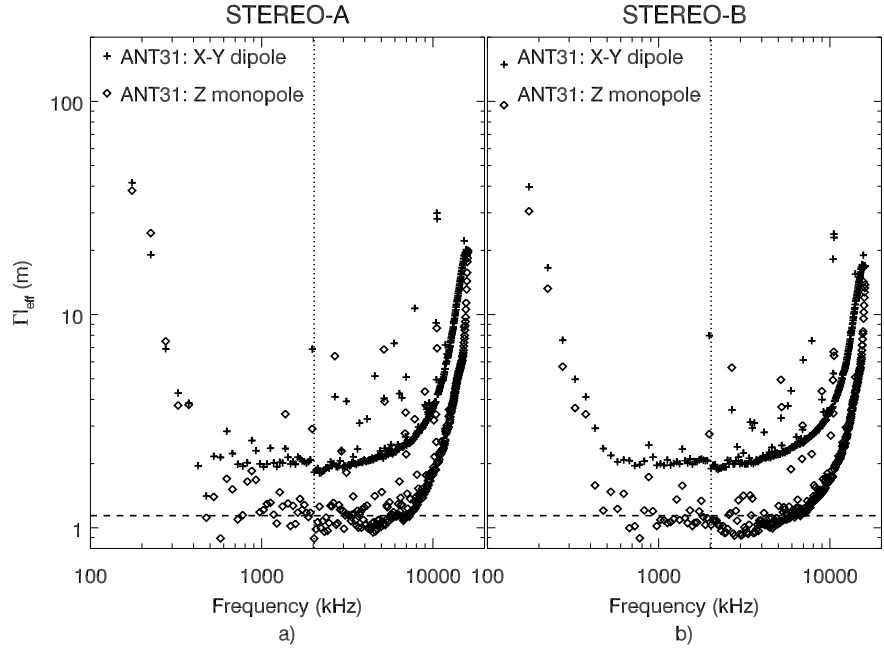}
              }
 \caption{Reduced effective antenna lengths of X--Y dipole (crosses) and Z monopole (diamonds)
for STEREO-A (left) and STEREO-B (right) vs. frequency obtained from measurements on January 13, 2007.
A black-dotted line separates frequency coverage of HFR1 ($<2$ MHz) and HFR2 ($>2$ MHz).
A dashed line indicates the $\Gamma l_{\rm{eff}}$ of the Z monopole
used for HFR1.}
   \label{sta31}
   \end{figure}

These values have been obtained by $\chi^2$-minimization considering the $V^2_{\rm{noise}}$ to be frequency independent \cite{2011RaSc...46.2008Z}.
We have performed the same analysis as \inlinecite{2011RaSc...46.2008Z} in order
to obtain accurate reduced effective antenna lengths
[$\Gamma l_{\rm{eff}}$]
as a function of frequency (Figure \ref{sta31}).	
These parameters have been retrieved by comparing the lowest $1$\,\% of the data observed within one day
(13 January 2007) and
the modeled galactic background [$V^2_{\rm{galaxy}}$]
using the Equation (\ref{eq:gammaleff}).
The low frequency part [$f<750$~kHz] is mostly affected by the shot noise 
induced by solar-wind particles impacting on the antennas \cite{2008SSRv..136..487B} and by the quasi-thermal noise (QTN) produced by the ambient plasma \cite{1989JGR....94.2405M}.
Since these effects are stronger than the galactic background, the method used
is not reliable for retrieving $\Gamma l_{\rm{eff}}$ for the low frequency part.

On the other hand, the short dipole approximation for STEREO antennas is not valid for frequencies above $\approx4$~MHz
since corresponding wavelengths become comparable with the reduced effective antenna length
[$\Gamma l_{\rm{eff}}$]
and thus induced voltages are not proportional to electric field fluctuations \cite{2008SSRv..136..529B}.
In Figure \ref{sta31} we can identify several frequency channels (24 for STEREO-A  and 26 for STEREO-B  from a total of 319, \textit{i.e.} $7$\,\%) with larger $\Gamma l_{\rm{eff}}$
than their neighboring frequency channels.
When the measured signal is significantly larger than the fit of the galactic background model,
it is likely that this discrepancy is due to instrumental effects, \textit{e.g.} frequency interferences \cite{2011RaSc...46.2008Z}.
We have  excluded these outlying values of $\Gamma l_{\rm{eff}}$ from our analysis.

For HFR1 ($f<2$~MHz) we have used fixed reduced antenna effective lengths (\textit{e.g.} see dashed lines in Figure \ref{sta31} for the Z antennas).
Measured fluxes by the three monopoles have been transformed into the orthogonal frame using effective antenna directions obtained
by observations of the non-thermal \textit{Auroral Kilometric Radiation} (AKR)
during STEREO-B  roll maneuvers \cite{2012JGRA..11706101K}.
As no AKR has been observed by STEREO-A, the effective directions have been assumed to be the same.
The three calculated orthogonal components of the flux density are summed to obtain the total flux density induced by an incident radio wave.

The antenna calibration for HFR2 ($f>2$~MHz) is more complex,
as two effects play a part, resulting in the reduced antenna effective
length being larger
than the physical one for higher frequencies above $\approx12$~MHz \cite{2011RaSc...46.2008Z}.
The first effect is the half-wave resonance that changes the effective antenna formulas to be functions of frequency.
The second one is the increase of $\Gamma$ above one
due to the electrical-circuit resonance as the antenna becomes inductive
($C_{\rm{a}} \rightarrow -C_{\rm{s}}$).
Above $12$~MHz the reduced effective antenna length rapidly changes and the conversion to physical units according to
Equation (\ref{eq:s}) becomes inaccurate.
For HFR2 we have used reduced antenna effective
lengths from Figure \ref{sta31} as a function of frequency. We have summed
both contributions of the X--Y~dipole and the Z-monopole to obtain the total flux density.

In order to improve the signal-to-noise ratio we have subtracted receiver background levels from the data before we performed our analysis.
These levels have been calculated as the median values over a period of 15 days (7 before, the corresponding one,
and 7 after) of the given auto-correlation for each channel--antenna configuration separately.

\section{Results and Discussion} 
      \label{S-Results}

We have manually selected 152 time--frequency intervals when Type III radio bursts have been observed by STEREO/\textit{Waves} between May 2007 and February 2013.
The separation angle between spacecraft in the ecliptic plane ranged between $7^\circ$ (May 2007) and $180^\circ$ (February 2011).
We have included only simple and isolated events when flux density was intense enough for the GP analysis.

\subsection{6 May 2009 Type III Radio Burst} 
      \label{S-event1}
\begin{figure}    
   \centerline{\includegraphics[width=0.9\textwidth,clip=,angle=0]{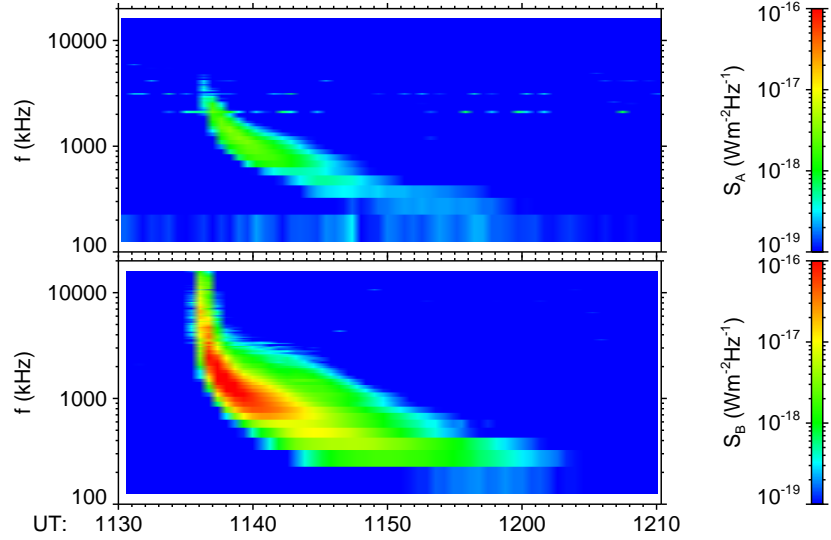}
              }
 \caption{Analysis of measurements recorded from 11:30 to 12:20 UT on 6 May 2009:
dynamic spectra for STEREO-A  (top)
and STEREO-B  (bottom).}
   \label{20090506_psd}
   \end{figure}

As an example from our list of events we present a Type III radio burst observed on
6 May 2009 by both STEREO spacecraft (Figure \ref{20090506_psd}).
A faint A3.8 X-ray flare located at N$20^\circ$E$65^\circ$ triggered the Type III radio burst.
During this event STEREO-A  was at $48.2^\circ$ West from
the Sun--Earth line at $0.96$~AU from the Sun whereas STEREO-B was located at $46.9^\circ$ East and $1.02$~AU from the Sun.
The separation angle between the STEREO spacecraft in the ecliptic plane was $95^\circ$.
STEREO-B, located near the source by $18.1^\circ$ longitude, observed
the Type III radio burst to have been much intense than STEREO-A being at $113.2^\circ$ longitude from the source.
Also the spectral shape of the emission at STEREO-B is much broader than at STEREO-A.
The onset time of the Type III radio burst was observed about one minute earlier at STEREO-B than at STEREO-A.
Whilst the frequency cutoff was not observed at STEREO-B, the radio burst is detected between $325$~kHz and $5$~MHz only at STEREO-A.

As we do not observe the triggering electron beam, we
are unable to use the method of \inlinecite{1980ApJ...236..696K} to distinguish the
\textit{F}- and \textit{H}-components of the Type III radio burst.
When both components are observed, the \textit{F}-component has typically larger flux density than the
\textit{H}-component \cite{1998JGR...10317223D}.
Using Monte Carlo simulations, \inlinecite{2007ApJ...671..894T} showed that scattering due to random density fluctuations at $120$~kHz extends the visibility of the \textit{F}-and \textit{H}-components
from $\approx18^\circ$ to $\approx90^\circ$ and from $\approx80^\circ$ to $\approx150^\circ$, respectively.
In other words, the \textit{F}-component is visible only in a narrow cone from a source only when compared to
\textit{H}-component being visible almost everywhere.
From spectral shapes and relative positions between the spacecraft and the flare site,
we may conclude that STEREO-B most probably observed the \textit{F}-
and \textit{H}-components whereas STEREO-A
measured the \textit{H}-component only.

\begin{figure}    
   \centerline{\includegraphics[width=0.7\textwidth,clip=,angle=90]{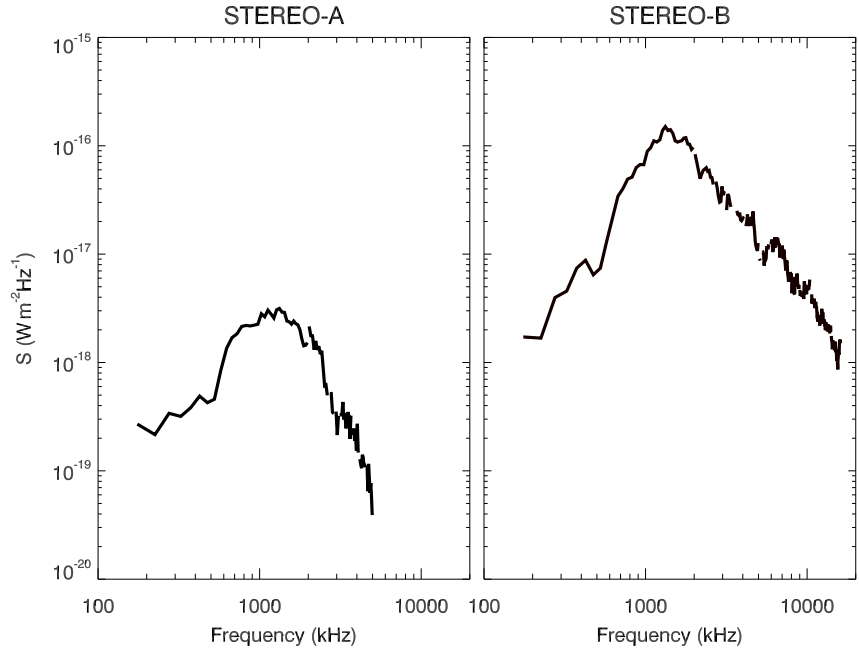}
              }
 \caption{Analysis of measurements recorded from 11:30 to 12:20 UT on 6 May 2009: flux density (peak flux values) vs.
frequency for STEREO-A  (left) and STEREO-B  (right).}
   \label{20090506_ls}
   \end{figure}

Figure \ref{20090506_ls} displays peak fluxes as a function of frequency for STEREO-A (on the left) and STEREO-B (on the right).
A signal detected at STEREO-B is about $100\times$ larger than at STEREO-A.
It is in agreement with position of the solar flare site, detected a lower signal at STEREO-A  (Figure \ref{20090506_ls}),
and an observed difference of onset times between the two spacecraft (Figure \ref{20090506_psd}).
The maximum flux density at both spacecraft occurs at $\approx1.5$ MHz.

We may identify a discontinuity in the HFR1/HFR2 connection at $2$~MHz on STEREO-A  but not on STEREO-B.

\subsection{23 November 2011 Type III Radio Burst} 
      \label{S-event2}
\begin{figure}    
   \centerline{\includegraphics[width=0.9\textwidth,clip=,angle=0]{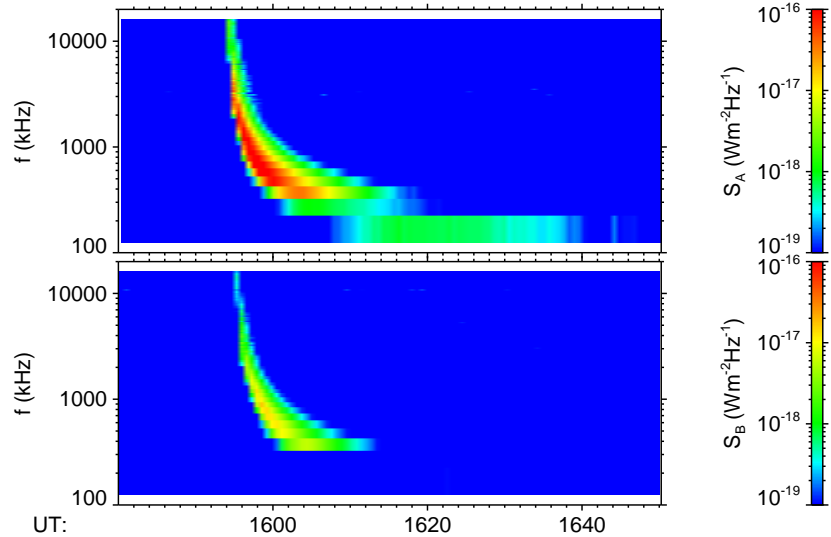}
              }
 \caption{Analysis of measurements recorded from 15:40 to 16:30 UT on 23 November 2011:
dynamic spectra for STEREO-A  (top) and STEREO-B
(bottom).}
   \label{20111123_psd}
   \end{figure}

As another example we present a Type III radio burst observed on
23 November 2011 by the two STEREO spacecraft.
STEREO-A  was at $106.1^\circ$ West from the
Sun--Earth line at $0.97$~AU from the Sun while STEREO-B was located $105.4^\circ$
East at $1.09$~AU from the Sun.
The separation angle between the STEREO spacecraft in the ecliptic plane was $149^\circ$.
Figure \ref{20111123_psd} shows flux density [$S$] from STEREO-A and STEREO-B  when an intense Type III radio burst
was detected at around 15:55 UT.
We have not linked this event to any solar flare as the active region was probably located on the far side of the Sun from the Earth's perspective.
The radio burst covers the full frequency range at STEREO-A. STEREO-B  observed the emission above $275$~kHz.

\begin{figure}    
   \centerline{\includegraphics[width=0.7\textwidth,clip=,angle=90]{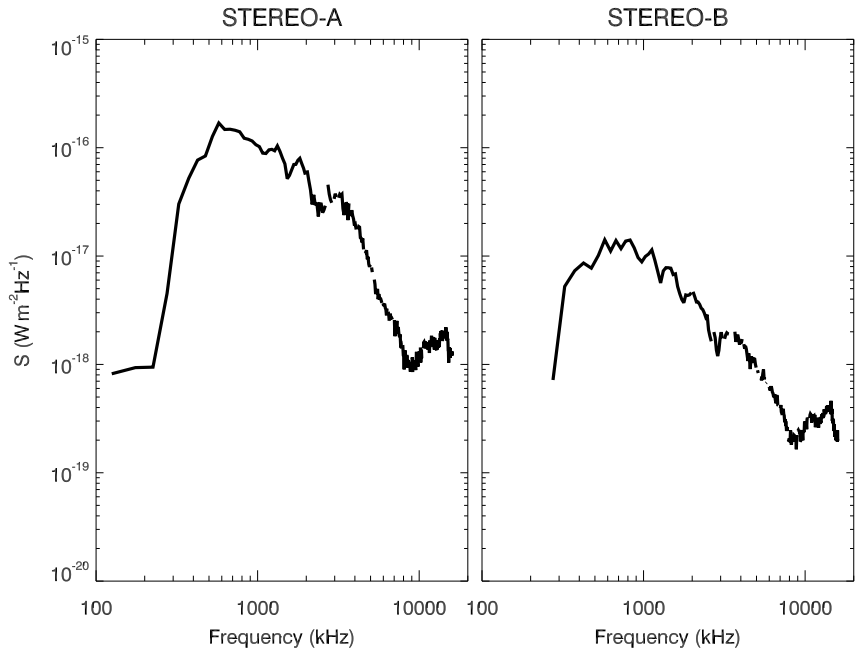}
              }
 \caption{Analysis of measurements recorded from 15:40 to 16:30 UT on 23 November 2011:
flux density (peak flux values) vs. frequency for STEREO-A  (left) and STEREO-B  (on the right).}
   \label{20111123_ls}
   \end{figure}

Figure \ref{20111123_ls} displays peak fluxes as a function of frequency for STEREO-A  (on the left) and STEREO-B  (on the right).
A signal detected at STEREO-A  is about ten times
larger than at STEREO-B. The maximum flux density occured at $625$~kHz.
We observed increase of the flux above $8$~MHz at both spacecraft. This increase can be caused by the antenna calibration when we are close to the half wave resonance.
Moreover, some additional calibration effects may take place here.
We have decided to keep such events in our data set since this flux increase might be a real feature of Type III radio bursts.

We do not identify any discontinuity in the HFR1/HFR2
connection at $2$~MHz, which suggests that we can combine
measurements of the monopole mode (HFR1) with the dipole/monopole mode (HFR2).

\subsection{Data Set} 
      \label{S-time}

Table \ref{tab:stat} summarizes Type III radio bursts included in our statistical survey.
Numbers of events from both spacecraft are about the same.
For the reasons described in Section~3.1 we do not distinguish \textit{F/H} components at long wavelengths in our data set. Generally, the \textit{F} component is more intense and more directive when compared to the \textit{H} component
\cite{2000GMS...119..115D,2007ApJ...671..894T}.
These two opposing effects cause uncertainty which of the two components is predominant in our data set.
Therefore, we will discuss our results considering that both plasma emission processes take place.

\begin{figure}    
   \centerline{\includegraphics[width=0.9\textwidth,clip=,angle=90]{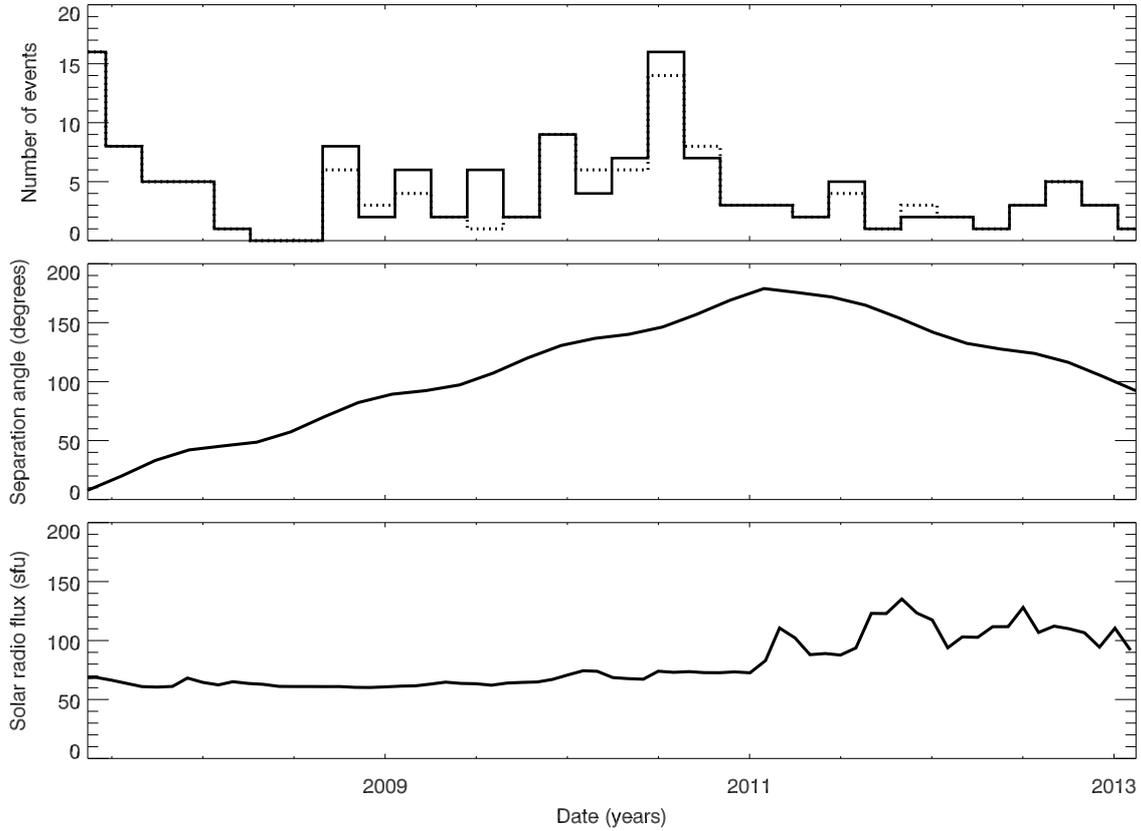}
              }
 \caption{Top panel: histogram of observed Type III radio bursts at STEREO-A  (solid line) and STEREO-B
(dotted line) between May 2007 and February 2013.
Middle: a separation angle between STEREO-A  and STEREO-B.
Bottom: monthly averages of the absolute value of
the solar radio flux ($1$~sfu~$\equiv10^{-22}\rm{Wm^{-2}Hz^{-1}}$) retrieved from \href{http://www.spaceweather.gc.ca/}{www.spaceweather.gc.ca}.}
   \label{histo_events}
   \end{figure}

The topmost panel of Figure \ref{histo_events} displays the histogram of the observed Type III radio bursts vs time
at STEREO-A  and STEREO-B. The middle panel is the separation
angle between STEREO-A  and STEREO-B.
The last panel contains absolute measurements of flux density from the Sun at a wavelength of 10.7 centimeters averaged over the month measured on the ground
(\href{http://www.spaceweather.gc.ca/}{\sf{www.spaceweather.gc.ca}}).
This value can be used as an estimator of the solar activity. Although the Sun exhibited increased activity as
of 2011 we do not have many events from this period since we
include only simple and isolated emissions.

We have compared onset times of Type III radio bursts with onset times of solar flares
listed
by the Lockheed Martin Solar and Astrophysics Laboratory
(\href{http://www.lmsal.com/solarsoft/latest\_events\_archive.html}{\sf{www.lmsal.com/solarsoft/latest\_events\_archive.html}}).
We have been able to link only $31$~bursts with solar flares (from total of 152, \textit{i.e.} $20$\,\% of all events).
Since X-ray imagers are located close to the Earth, we have more linked events in 2007 (up to $45$\,\%,
when the two STEREO were near the Earth)
compared to 2012\,--\,2013 when only few Type III radio bursts have been associated (the two STEREO spacecraft were on the far side of the Sun).

\begin{figure}    
   \centerline{\includegraphics[width=0.4\textwidth,clip=,angle=90]{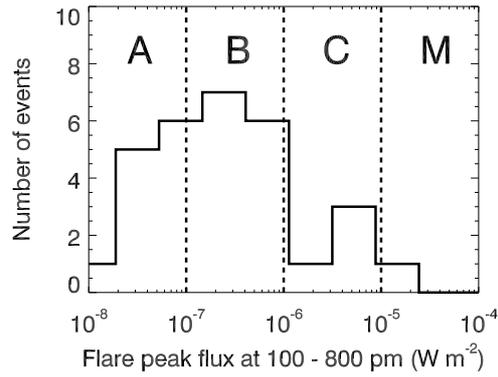}
              }
 \caption{Histogram of solar flares associated with Type III radio bursts.}
   \label{flare_histo}
   \end{figure}

Figure \ref{flare_histo} shows a histogram of solar flares responsible for the aforementioned 31 Type III radio bursts
(flare class A: 6 events, flare class B: 15 events, flare class C: 9 events, flare class M: 1 event).
Generally, more intense solar flares trigger complex Type III radio bursts which are not
the area of our interest.

We have also compared events from our data set with ground-based observations. The results
will be addressed in a separate publication.

\subsection{Frequency Distribution} 
      \label{S-frq}
\begin{figure}    
   \centerline{\includegraphics[width=0.4\textwidth,clip=,angle=90]{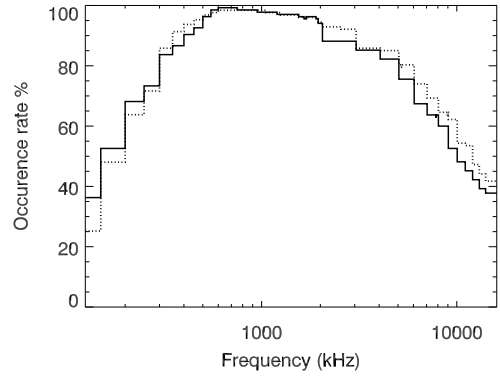}
              }
 \caption{The histogram of frequencies of observed Type III radio bursts at STEREO-A
(solid line) and STEREO-B (dotted line).}
   \label{frq_dist}
   \end{figure}

Figure \ref{frq_dist} displays histograms of frequencies of all Type III radio bursts from our data set for STEREO-A  (solid line) and STEREO-B (dotted line).
When combined with the number of events (Table~1) it demonstrates the number of data points used in our survey.
Although $f_{\rm{pe}}$ at STEREO is typically around $30$~kHz, only $35$\,\% (STEREO-A) and $25$\,\% (STEREO-B) of events have been measured up to $125$~kHz corresponding to the lowest-frequency channel of
HFR.
Our statistical results on low-frequency cutoffs are generally comparable with 
previous studies dedicated to Wind and Ulysses observations, \textsl{e.g.} by \inlinecite{1996A&A...316..406L} and \inlinecite{1996GeoRL..23.1203D}.
\inlinecite{1995GeoRL..22.3429L} suggest that the low-frequency cutoff can be explained as
i) an intrinsic characteristic of the radiation  mechanism,
ii) an effect of a directivity of the radiation, and iii)
propagation effects between the source and the observer.

On the other hand more than $40$\,\% of Type III radio bursts are observed up to $16$~MHz, which corresponds to a source
position located $\approx1$ $\rm{R}_{\odot}$ and $\approx1.6$ $\rm{R}_{\odot}$ above
the Sun's surface assuming the \textit{F}- and \textit{H}-emission,
respectively (Figure \ref{sittler}).
Concerning high-frequency cutoffs, \inlinecite{2000GMS...119..115D} investigated $269$~events observed by the \textit{Wind} spacecraft and
$70$\,\% of them started at frequencies below $13.8$~MHz \--- the maximum frequency recorded by the instrument.
In our dataset we have around $50$\,\%
of radio bursts that commence at frequencies below $13$~MHz.


\subsection{Statistical Analysis of the Frequency Spectra} 
      \label{S-stat}

We have identified frequencies corresponding to the maximum-flux density
for each Type III radio burst separately.
Table \ref{tab:max_frq} contains statistical distributions of these peak frequencies.
Medians of peak frequencies are around $1$~MHz ($825$~kHz at STEREO-A  and $925$~kHz at STEREO-A) which is in agreement with previous observations.
\inlinecite{1978SoPh...59..377W} found in data from the \textit{Interplanetary Monitoring Platform-H} (IMP-6) spacecraft ($30-9900$ kHz) that the maximum flux density occurs at $\approx 1$ MHz.

We have investigated median values of the flux density $S$ vs. frequency (see Figure \ref{int1}).
As the distribution has a log-normal character, we have used median values instead of
mean ones. The maximum flux density ($3\times10^{-18}$ $\rm{Wm^{-2}Hz^{-1}}$ or $3\times10^{4}$ sfu) occurs at $\approx 1$ MHz on both spacecraft.
The low-frequency part is affected by the QTN and shot noise, especially for STEREO-A.
We have observed a slight increase of flux at STEREO-A  for frequencies above $10$~MHz but not at STEREO-B.
In the case of STEREO-B,  it is covered by noise as the
$75$th percentile exhibits the same trend (Figures \ref{20090506_ls} and \ref{20111123_ls}).
This increase can be caused by inconsistencies of the antenna calibration when we are close to the half-wave resonance (Figure \ref{sta31}).
On the other hand it can be a real feature of Type III radio bursts. Unfortunately, we
cannot compare our results with previous studies since this frequency range ($8$~MHz --- $16$~MHz) has not been
investigated by space-borne nor ground-based instruments.
Therefore we conclude that measurements of frequencies above $8$~MHz should be considered carefully.

\subsection{Interpretation of the $1$~MHz maximum} 
  \label{S-model}
	
\begin{figure}    
   \centerline{\includegraphics[width=0.6\textwidth,clip=,angle=90]{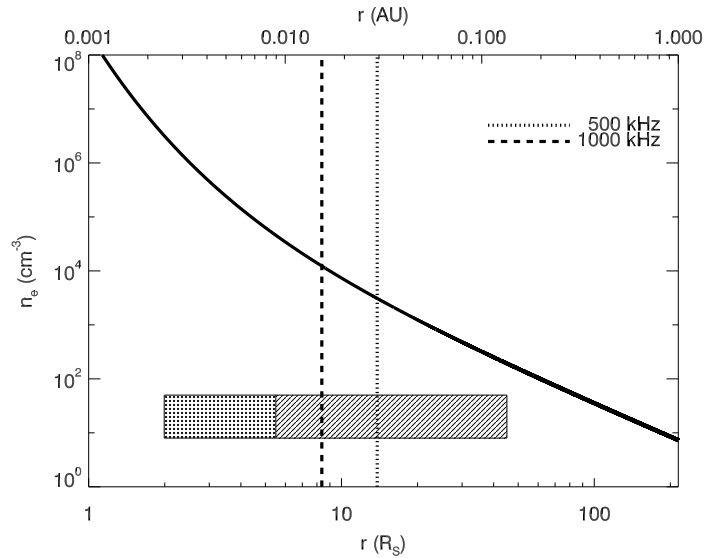}
              }
 \caption{Electron density model vs. distance from the Sun for average solar wind parameters adapted from Sittler and Guhathakurta (1999).
The dashed region indicates frequencies where the STEREO/\textit{Waves}/HFR provides us with the GP data while
in the dotted region we can retrieve information on wave intensity only. Dashed
and dotted lines indicate radial distances
from the Sun where electron density corresponds to a plasma frequency of $1$~MHz and $500$~kHz, respectively.}
   \label{sittler}
   \end{figure}

\begin{figure}    
   \centerline{\includegraphics[width=0.7\textwidth,clip=,angle=90]{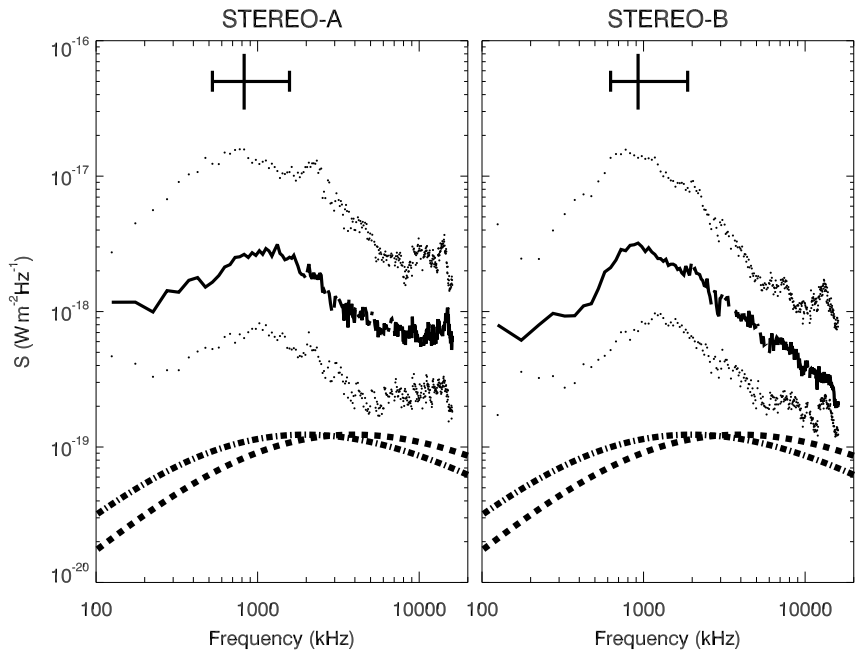}
              }
 \caption{Flux density vs. frequency at STEREO-A  and STEREO-B 
for 152 Type III radio bursts.
Solid lines are the medians of the flux for each frequency and dotted lines represent $25$th and $75$th percentiles.
Black crosses on the top denote median and $25$th/$75$th percentiles of the peak frequency from Table \ref{tab:max_frq}.
Dashed and dashed--dotted lines represent the model of the
\textit{H}- and \textit{F}-component in arbitrary units, respectively.}
   \label{int1}
   \end{figure}

Using the electron-density model of the solar wind we can determine average distances of radio sources from the Sun of Type III radio bursts (Figure \ref{sittler}).
A frequency of $1$ MHz corresponds to a radio source located at $\approx8$~$\rm{R}_{\odot}$ and
$\approx14$~$\rm{R}_{\odot}$ from the Sun for the \textit{F}- and \textit{H}-component, respectively (dashed and dotted lines in Figure \ref{sittler}).
The plasma density [$n(r)$] in the corona decreases faster than $r^{-2}$, but starting from around $7$~$\rm{R}_{\odot}$ it decreases as $r^{-2}$
(a solid line in Figure \ref{sittler}).
The $1$~MHz maximum in Figure \ref{int1} coincides with this region.
One should note that the critical radius of Parker's model of the solar-wind expansion is typically about $10$~$\rm{R}_{\odot}$
being roughly between $8$~$\rm{R}_{\odot}$ and $14$~$\rm{R}_{\odot}$.

We have developed a simplified analytical model of the flux density as a function of radial distance.
Let us consider propagation and generation of plasma waves in the corona. Assuming fast relaxation of the beam,
we find that the beam propagates
as a beam-plasma structure (e.g. \opencite{2001A&A...375..629K}), so that the 1D electron distribution function [$f(x,v,t)$] is
   \begin{eqnarray}
     f (x,v,t) &=&\frac{n_{\rm{b}}}{n_0}\exp{\left(-\frac{(x-v_0 t/2)^2}{d^2}\right)}, v<v_0
                       \nonumber \\      
                 &=& 0, v>v_0\,,
                       \label{eq:f1}    
   \end{eqnarray}
and the spectral energy density of Langmuir waves with wavenumber $k=\omega_{pe}/v$ is
   \begin{equation}  \label{eq:w1}
     W (x,v,t) = \frac{m_{\rm{e}} n_{\rm{b}}}{v_0 \omega_{\rm{pe}}} v^4 \left(1-\frac{v}{v_0}\right)\exp{\left(-\frac{(x-v_0 t/2)^2}{d^2}\right)}, v<v_0\,,
   \end{equation}
where $f (x,v,t)$ is the plateau-like electron distribution function of the beam electrons, $n_{\rm{b}}$ their number density, $v_0$ is
the maximum velocity of the plateau electrons,
$m_{\rm{e}}$ is electron mass, and $W (x,v,t)$ is the 1D Langmuir-Waves spectral density.
Let us first consider emission at the spatial peak, e.g. $x \approx v_0t/2$ and near the spectral peak $v \approx 4/5v_0$:
   \begin{equation}  \label{eq:w2}
     W (x,v,t) \simeq \frac{n_{\rm{b}}}{\omega_{pe}} v_0^3 \,.
   \end{equation}
In a case where of the \textsl{H}-component
of an electromagnetic (EM) radiation is saturated we
consider its spectral energy density [$W_H^{\rm{EM}}$]
to be proportional to the energy
of Langmuir waves \cite{1980SSRv...26....3M}:
   \begin{equation}  \label{eq:w2}
     W_H^{\rm{EM}} (x,v,t) \simeq W(x,v,t)\,.
   \end{equation}

The spectral flux density of the \textit{H}-component
at the Earth [$S_H$] can
be estimated (the same as if the
spacecraft is at $1$~AU):

   \begin{equation}  \label{eq:s1}
     S_H = W_H^{\rm{EM}} \frac{\mbox{d} k}{\mbox{d} \omega}\,A v_{\rm{g}} \frac{1}{4 \pi R_{\phi}^2}\,,
   \end{equation}

where $\mbox{d} k / \mbox{d}\omega = 1/v_{\rm{g}}$, $v_{\rm{g}}$ is the group velocity of the EM waves,
$A$ is the area of the source, and $R_{\phi}$ is the $1$~AU distance.
Hence one finds, ignoring constants:

\begin{equation}  \label{eq:s2}
     S_H \propto \frac{n_{\rm{b}}}{ f_{\rm{pe}}} v_0^3 A\,,
   \end{equation}
where $f_{\rm{pe}}$ is the plasma frequency. Further assuming that $n_{\rm{b}} Av_0^3 \propto r^{-2}$,
one obtains

   \begin{equation}  \label{eq:s4}
     S_H \propto \frac{1}{r^2} \frac{1}{f_{\rm{pe}}}\,.
   \end{equation}
We have included the model in arbitrary units in Figure \ref{int1}.
The semiempirical model of electron density of \inlinecite{1999ApJ...523..812S} in the solar corona and IP medium has been used for estimating $f_{\rm{pe}}$.
The model of the flux density as a function of frequency [$S_H(f=2f_{\rm{pe}})$]
exhibits a maximum at $4$~MHz
(corresponding to a radio source located at $5$~$\rm{R}_{\odot}$ from the Sun's center)
that differs from the observation (maximum at $1$~MHz).
The model assumes the \textit{H}-emission only since a relation between the spectral energy density
for the \textit{F}-emission $W_F^{\rm{EM}}$ and the energy of Langmuir waves
[$W$] is not obvious.

However, we have modified the model
for the \textsl{F}-component considering that the radio flux is proportional between the two components [$S_F(f=f_{\rm{pe}})$].
In this case we have obtained a maximum at $2$~MHz being in a better
agreement with the observed maximum at $1$~MHz.
One should note that this approach is very simplified omitting many physical processes such as an efficiency
of Langmuir and radio waves conversion, volume of source regions, and scattering of radio beams by density fluctuations.

%
%
\section{Summary and Concluding Remarks} 
      \label{S-Conclusion}
Type III radio bursts belong among the most intense electromagnetic emissions in the
heliosphere. They can be observed from tens of kHz up to several GHz. They are frequently detected by STEREO/\textit{Waves} instruments, which
provide us with a unique stereoscopic view of radio sources in a frequency range from $125$~kHz to $16$~MHz. These frequencies correspond to
radial distances from the Sun from $1$~$\rm{R}_{\odot}$ to $50$~$\rm{R}_{\odot}$ and $1.6$~$\rm{R}_{\odot}$ to $85$~$\rm{R}_{\odot}$
for the \textit{F}- and \textit{H}-component, respectively.

We have investigated properties of $152$ simple and isolated Type III radio bursts observed by the two STEREO spacecraft
between May 2007 and February 2013.
We show a detailed analysis of two events from our data set, as an example.

Our main statistical results on the frequency spectra are as follows:
\begin{enumerate}
	\item[i)] We have observed the low-frequency cutoff at $125$~kHz in $65$\,\% --- $75$\,\% of events.
	\item[ii)] More than $40$\,\% of Type III radio bursts are observed up to $16$~MHz, which corresponds to a source position located
	$\approx1$~R$_{\odot}$ and $\approx1.6$~R$_{\odot}$	from the Sun's surface
	for the \textit{F}- and \textit{H}-component, respectively.
	\item[iii)] We have found that the maximum of flux density
	($S \approx 3\times10^{-18}$ $\rm{Wm^{-2}Hz^{-1}}$) occurs at $1$~MHz on both spacecraft which corresponds to
	radio sources located at $8$~$\rm{R}_{\odot}$ and $14$~$\rm{R}_{\odot}$ for the
	\textit{F}- and \textit{H}-component, respectively.
	\item[iv)] We have introduced a simple model of the radio emission flux density vs. radial distance which provides us with a peak at a frequency within a factor of
four (the \textit{H}-component) and two (the \textit{F}-component)
from the observation, respectively.
\end{enumerate}

The observed low- high-frequency
cutoffs can be explained by intrinsic features of radio sources, their directivity,  
and/or propagation effects between the source and the spacecraft.
The source region of radio emissions at $1$~MHz corresponds to the region where a change of the plasma density gradient takes place.

We have also observed an increase of the flux density above $8$~MHz, which corresponds to radial distances from the Sun of
 $1.6$~$\rm{R}_{\odot}$ and $2.6$~$\rm{R}_{\odot}$ for the \textit{F}- and
\textit{H}-component, respectively.
We conclude that results of the radio flux above $8$~MHz are questionable probably due to instrumental effects.

The model for the \textit{H}-component exhibits a peak at $4$~MHz.
We have modified the model for the \textit{F}-component
and the obtained maximum at $2$~MHz better corresponds to the maximum observed in the data.
One should note that various assumptions on an electron density profile in the
solar wind, variations of an exciter speed, and an electron beam density profile may lead to different maximum locations.

In the future we plan to perform simulations
of the resonant interaction of an electron beam with Langmuir waves 
in connection with the observed $1$~MHz maximum.

\section*{Acknowledgements} 
      \label{S-ack}
The authors would like to thank the many individuals and institutions who contributed to making
STEREO/\textit{Waves} possible.
J.~Soucek acknowledges the support of the Czech Grant
Agency grant GAP209/12/2394.
O.~Kruparova thanks the support of the Czech Grant
Agency grant GP13-37174P.
Financial support by STFC and by the European Commission through
the ''Radiosun'' (PEOPLE-2011-IRSES-295272) is gratefully acknowledged (E.P.~Kontar). 
O.~Santolik and V.~Krupar acknowledge the support of the Czech Grant
Agency grant GAP205/10/2279.

\begin{table}[htb]
\begin{tabular}{lr}
    \hline
STEREO-A  events & 135 \\
STEREO-B  events & 127 \\
\hline
STEREO-A  and STEREO-B  & 110 \\
STEREO-A  and not STEREO-B & 25 \\
STEREO-B  and not STEREO-A  & 17 \\
\hline
\textbf{Total number of events}& \textbf{152}\\
    \hline
\end{tabular}
\caption{The number of radio bursts detected at STEREO-A and at STEREO-B  are in the first and second row, respectively. The third row contains a count of simultaneously observed events.
The fourth and fifth rows show events that have been observed only by one spacecraft. The total number of Type III radio bursts in our survey is in the last row.}
\label{tab:stat}
\end{table}

\begin{table}[htb]
\centering
\begin{tabular}{lrr}
\hline
~  & STEREO-A  & STEREO-B  \\
\hline\\
Mean [kHz]  & 1291 & 1498 \\
STD [kHz]   & 1534 & 1894 \\
Median [kHz] & 825  & 925  \\
$25$th percentile [kHz]   & 525  & 625  \\
$75$th percentile [kHz]  &1575  & 1875  \\
    \hline
\end{tabular}
\caption{Statistical properties of peak frequencies of Type III radio bursts included in the survey.}
\label{tab:max_frq}
\end{table}

%
\bibliographystyle{spr-mp-sola}
%
\bibliography{references}

\end{article}

\end{document}